\documentclass[useAMS,usenatbib]{mn2e}
\usepackage{amsmath}
\usepackage{usegraphicx}

\newcommand{\myemail}{enrique.lopezrodriguez@utsa.edu}
\newcommand{\degree}{$^{\circ}$}
\newcommand{\um}{$\mu$m}

\title[The torus of NGC 1068 as a hydromagnetic outflow wind]{Near-Infrared Polarimetric Adaptive Optics Observations of NGC 1068: A torus created by a hydromagnetic outflow wind}
\author[E. Lopez-Rodriguez et al.]{E. Lopez-Rodriguez$^{1}$\thanks{E-mail: \myemail}\thanks{Researcher Affiliated-Postdoctoral, Department of Astronomy, University of Texas at Austin}, C. Packham$^{1}$, T. J. Jones$^{2}$, R. Nikutta$^{3}$, L. McMaster$^{1}$ \newauthor
R.~E.~Mason$^{4}$, M. Elvis$^{5}$, D. Shenoy$^{2}$, A. Alonso-Herrero$^{6}$\thanks{Visiting Professor, Department of Physics and Astronomy, University of
Texas at San Antonio}, E. Ram\'irez$^{7}$, \newauthor
O. Gonz\'alez Mart\'in$^{8,9,10}$, S.~F. H\"onig$^{11}$,  N.~A.~Levenson$^{12}$, C. Ramos Almeida$^{9,10}$\thanks{Marie Curie Fellow}, \newauthor
E. Perlman$^{13}$ \\
	$^{1}$Department of Physics \& Astronomy, University of Texas at San Antonio, One UTSA Circle, San Antonio, TX 78249, USA \\
	$^{2}$Minnesota Institute for Astrophysics, University of Minnesota, 116 Church Street SE, Minneapolis, MN 55455, USA \\
	$^{3}$Instituto de Astrof\'isica, Facultad de F\'isica, Pontificia Universidad Cat\'olica de Chile, 306, Santiago 22, Chile \\
	$^{4}$Gemini Observatory, Northern Operations Center, 670 N. A'ohoku Place, Hilo, HI 96720, USA \\
	$^{5}$Harvard Smithsonian Center for Astrophysics, 60 Garden St., Cambridge, MA 02138, USA \\	
	$^{6}$Instituto de F\'isica de Cantabria, CSIC-UC, E-39005 Santander, Spain \\
	$^{7}$Universidade de S\~ao Paulo, IAG, Rua do Mat\~ao 1226, Cidade Universit\'aria, S\~ao Paulo 05508-900, Brazil. \\
	$^{8}$Centro de Radioastronom\'ia y Astrof\'isica (CRyA-UNAM), 3-72 (Xangari), 8701, Morelia, Mexico \\
	$^{9}$Instituto de Astrof\'isica de Canarias, Calle V\'ia L\'actea s/n, 38205, Tenerife, Spain \\
	$^{10}$Universidad de La Laguna, Departamento de Astrof\'isica, E-38206 La Laguna, Tenerife, Spain \\
	$^{11}$School of Physics \& Astronomy, University of Southampton, Southampton, SO17 1BJ, United Kingdom \\
	$^{12}$Gemini Observatory, Casilla 603, La Serena, Chile \\
	$^{13}$Department of Physics and Space Sciences, Florida Institute of Technology, Melbourne, FL 32901
	}
	
\begin{document}

\date{Accepted XXXX. Received XXXX; in original form XXXX. ACCEPTED by MNRAS}

\pagerange{\pageref{firstpage}--\pageref{lastpage}} \pubyear{2014}

\maketitle

\label{firstpage}

\begin{abstract}
We present J$'$ and K$'$ imaging linear polarimetric adaptive optics observations of NGC 1068 using MMT-Pol on the 6.5-m MMT. These observations allow us to study the torus from a magnetohydrodynamical (MHD) framework. In a 0$\farcs$5 (30 pc) aperture at K$'$, we find that polarisation arising from the passage of radiation from the inner edge of the torus through magnetically aligned dust grains in the clumps is the dominant polarisation mechanism, with an intrinsic polarisation of 7.0\%$\pm$2.2\%. This result yields a torus magnetic field strength in the range of 4$-$82 mG through paramagnetic alignment, and 139$^{+11}_{-20}$ mG through the Chandrasekhar-Fermi method. The measured position angle (P.A.) of polarisation at K$'$ is found to be similar to the P.A. of the obscuring dusty component at few parsec scales using infrared interferometric techniques. We show that the constant component of the magnetic field is responsible for the alignment of the dust grains, and aligned with the torus axis onto the plane of the sky. Adopting this magnetic field configuration and the physical conditions of the clumps in the MHD outflow wind model, we estimate a mass outflow rate $\le$0.17 M$_{\odot}$ yr$^{-1}$ at 0.4 pc from the central engine for those clumps showing near-infrared dichroism. The models used were able to create the torus in a timescale of $\geq$10$^{5}$ yr with a rotational velocity of $\leq$1228 km s$^{-1}$ at 0.4 pc. We conclude that the evolution, morphology and kinematics of the torus in NGC 1068 can be explained within a MHD framework.
\end{abstract}


\section{Introduction}
\label{INTRO}

The torus is the cornerstone of the unified model \citep{L1991,A1993,UP1995} of Active Galactic Nuclei (AGNs). Specifically, this model posits that the observational differences between radio-quiet AGNs arise from an orientation effect. In this scheme, the AGN classification solely depends on the anisotropic obscuration of the central engine (black hole and accretion disc) by an optically and geometrically thick, dusty torus. Although the torus was postulated to be a ``donut-shape'' morphology, in this paper the term ``torus'' is used to denote the region surrounding the central engine where the obscuring material is located, with the precise morphology of that region still to be determined.

Several ideas to explain the existence of the torus have been proposed. Some models explained the torus as an inflow of gas from few tens of kpc. \citet{WPS2009} presented numerical simulations of the interstellar medium to track the formation of molecular hydrogen forming an inhomogeneous thick disc around the central engine, identified as the torus. In a further 3D hydrodynamic simulation model, \citet{W2012} found that a vertical circulation of gas in the central parsecs drives the formation of an obscuring region along the equatorial plane. \citet{SKB2011} suggested the origin of the torus as dusty clouds falling from the host galaxy to the central engine. Conversely, some models \citep{BP1982,KB1988,EBS1992,KK1994,K1999,ES2006} assume an outflowing wind confined and accelerated by the magnetic field generated in the accretion disc. In this scheme, the hydromagnetic wind can lift the plasma from the midplane of the accretion disc to form a geometrically thick distribution of dusty clouds surrounding the central engine. High angular resolution observations suggest the infrared (IR) emitting regions of the torus to no more that a few parsecs in radius \citep[i.e.,][]{Radomski2003,Jaffe2004,Packham2005,Mason2006,Radomski2008,RA09,Raban2009,RA11,AH2011,B2013,I2014,MKN2014}. Within these scales, the torus is in the region where the supermassive black hole (SMBH) and accretion disc activity dominate over the galaxy components. Thus, the torus origin is likely influenced by the central engine, which gives strong impetus to the outflowing wind model. 

Although large efforts have been made in the development of (magneto-)hydrodynamical (MHD) simulations of the outflow wind of AGNs, the magnetic field strengths in the torus are poorly constrained. The magnetic field can induce a preferential orientation of dust grains within the torus that can give rise to a measurable degree of polarisation \citep[e.g.][]{DG1951,JS1967,JKD1992}, with an expected linear degree of polarisation $<$15\% at near-IR (NIR: 1$-$5 \um) assuming standard disc models \citep{EMH1997}. Hence, IR polarimetric techniques give us a powerful tool to enhance the contrast of the polarised structures arising from aligned dust grains, helping us to constrain the magnetic fields, from those unpolarised components within the core of AGNs. Using J, H and K$_{\mbox{\tiny n}}$ imaging polarimetric observations on the 3.9-m Anglo-Australian Telescope, \citet{LR2013} estimated a magnetic field strength in the range of 12$-$128 mG in the NIR emitting regions in the torus of IC 5063. Through comparison with the magnetic field strength of $\sim$2 mG estimated by the ratio of thermal and magnetic field pressure, they suggested that a magnetically dominant region is surrounding the central engine of IC 5063. They concluded that the magnetic field plays an important role in the torus evolution of IC 5063, favoring the MHD outflow wind model.  

NGC 1068 is an archetypal type 2 AGN, whose proximity (we adopt H$_{0}$ = 73 km s$^{-1}$ kpc$^{-1}$, thus 1$''$ = 60 pc) and high brightness make it an ideal target for polarimetry. The detection of polarised broad emission lines in the optical (0.35$-$0.70 \um) wavelengths of NGC 1068 by \citet{AM85} was crucial to understand the AGN structure. Their detection is most readily interpreted through scattering of the radiation from the central engine into our line of sight (LOS) by the ionisation cones. After correction for starlight dilution, the continuum polarisation was shown to be constant, $\sim$16\%, over the ultraviolet \citep{AHM94} and optical wavelengths \citep{MA83}, with electron scattering being the dominant polarisation mechanism. Further optical (0.5$-$0.6 \um) polarimetric studies \citep{Capetti1995} using the {\it Hubble Space Telescope} ({\it HST}) showed a centrosymetric polarisation pattern along the ionisation cones in the $\sim$10$''$ (600 pc) central region. This polarisation pattern is the signature of a central point source whose radiation (which can be polarised from multiple scattering into the funnel of the torus and/or broad line clouds)  is ultimately polarised through scattering by dust and/or electrons. At ultraviolet and optical wavelengths, these studies showed that electron scattering is the dominant polarisation mechanism in the core and ionisation cones of NGC1068. In the NIR, an increase in the polarisation flux density with increasing wavelength in the core was observed \citep[e.g.][]{LKR78,Brindle1990}. Further modeling using NIR polarimetric observations has shown that the polarisation in the core most likely arises from dichroic absorption of radiation from the central engine by aligned dust grains in the torus \citep{Young1995,Packham1997,Lumsden1999,Simpson2002,Watanabe2003}. A combination of electron and dust scattering from the ionisation cones still has some contribution, i.e. $<$10\% of the polarised flux at 2.0 \um~\citep{Young1995}. \citet{HPB2008} estimated that the contribution of the accretion disc emission to the total flux is negligible at 2.0 \um, and that dust emission from a clumpy torus is the dominant contributor. In addition, \citet{Watanabe2003} raise the possibility that scattering off large grains in the torus could also explain the NIR polarisation. They argued that the change of position angle (P.A.) of polarisation and the increase in the degree of polarisation from optical to NIR due to dust scattering by large grains in the torus cannot be ruled out. Although detailed studies have not been done yet, some implications are discussed in Section \ref{ANA_INT_POL}. 

In the present paper, we examine the MHD outflow wind model in the core of NGC 1068 assuming the torus to be clumpy \citep{Nenkova2002,Nenkova2008a,Nenkova2008b}. We follow the approach from \citet{LR2013} to estimate the magnetic field strength in the torus of NGC 1068 and extent the study to estimate physical parameters of the torus using MHD outflow wind models, allowing us to describe the evolution and kinematics of the clumps. We performed imaging polarimetric observations in the J$'$ and K$'$ filters using MMT-Pol in conjunction with the adaptive optics (AO) system on the 6.5-m MMT. The paper is organised as follows: Section \ref{OBS} describes the observations and data reduction, Section \ref{RES} presents our polarimetric results, which are analyzed in Section \ref{ANA}. In Section \ref{MAG} we estimate the magnetic field strength in the torus of NGC 1068. Section \ref{DIS} presents the discussion and Section \ref{CON} the conclusions.




\section{Observations and data reduction}
\label{OBS}

NGC 1068 was observed on 2013 October 23 using MMT-Pol \citep{PJ2008,P2010,P2012} in conjunction with the AO system and the f/15 camera on the 6.5-m MMT, Arizona. The AO system was used to assist the polarimetric observations of MMT-Pol. Specifically, a dichroic at 15\degree~to the normal before the aperture window of MMT-Pol reflects optical light up to a CCD-based wavefront sensor and passes the IR beam into MMT-Pol at the Cassegrain focus. Thus, NGC 1068 is used for the AO correction at optical wavelengths. MMT-Pol uses a 1024 $\times$ 1024 pixels HgCdTe AR-coated Virgo array, with a pixel scale of 0$\farcs$043 pixel$^{-1}$, corresponding to a field of view (FOV) of 44$''$ $\times$ 44$''$. MMT-Pol uses a rectangular focal plane aperture, a half-wave retarder (half wave plate, HWP), and one of two Wollaston prisms. The rectangular focal plane aperture provides two non-overlapping rectangular images with an individual FOV of 20$'' \times$ 40$''$. In standard polarimetric observations, the HWP is rotated to four P.A. in the following sequence: 0\degree, 45\degree, 22$\fdg$5, and 67$\fdg$5. A Calcite Wollaston is used in the wavelength range of 1$-$2 \um~and a Rutile Wollaston is used in the wavelength range of 2$-$5 \um.

The J$'$ ($\lambda_{\mbox{\tiny c}}$=1.33 \um, $\Delta\lambda$=0.08 \um, 50\% cut-on/off) and K$'$ ($\lambda_{\mbox{\tiny c}}$=2.20 \um, $\Delta\lambda$=0.11 \um, 50\% cut-on/off) filters provide the best combination of sensitivity and wavelength range within the instrumental filter set; thus, these filters were used for the observations. In both filters, the images were acquired in an ABA dither pattern with an offset of 8$''$ in declination, where four HWP P.A. were taken in each dither position. The position of the short axis of the array with the north on the sky was 149\degree~and 46\degree~E of N at J$'$ and K$'$, respectively. At J$'$, frame exposure times of 15s per HWP P.A. at each dither position were taken, with a total of 3 ABA dither patterns, providing a total exposure time of 540s. At K$'$, frame exposure times of 10s per HWP at each dither position were taken, with a total of 10 ABA dither patterns, providing a total exposure of 1200s. A summary of observations is shown in Table \ref{Table1}.

The data were reduced using custom \texttt{IDL} routines and standard NIR imaging procedures. The difference for each correlated double sample (CDS) pair was calculated, and then sky subtracted using the closest dither position in time to create a single image per HWP P.A. For each dither position, the ordinary (o-ray) and extraordinary (e-ray) rays, produced by the Wollaston prism, were extracted and then the Stokes parameters, $I$, $Q$, and $U$ were estimated according to the ratio method \citep[e.g.][]{Tinbergen2006} for each dither position. The Stokes parameters were registered and shifted to a common position, and then co-averaged to obtain the final $I$, $Q$, and $U$ images. Finally, the degree, $P = \sqrt{Q^2+U^2}/I$, and P.A., $PA = 0.5\arctan(U/Q)$, of polarisation were estimated. During this process, individual photometric and polarimetric measurements were performed for each dither position, allowing to examine for a high and/or variable background that could indicate the presence of clouds or electronic problems. Fortunately, no data needed to be removed for these reasons, but some data were removed when the AO guide unlocked. The measurements of the degree of polarisation were corrected for polarisation bias using the approach by \citet{WK1974}.

The image of the unpolarised standard star,  HD 224467, in the  K$'$ filter was used to estimate the image quality of the observations, ensuring the quality of the AO system prior the science observations. For the standard star, the full width at half maximum (FWHM) using a Gaussian profile was estimated to be 0$\farcs$17 $\times$ 0$\farcs$23 in the K$'$ filter (Figure \ref{fig1}, bottom-middle). MMT-Pol has a residual astigmatism that produces an X-shape on the FWHM. The polarised standard star, HD 38563C, was observed in both filters. The observations were performed during windy conditions on the observatory. The unpolarised standard star, HD 224467, was observed against the wind, minimizing instrumental shakes, whilst NGC 1068 observations were performed in favor of the wind, which affect the image quality. Frame exposure times of 10s per HWP P.A. at each dither position were taken, with a total of 2 and 3 ABA dither patterns, providing a total exposure time of 240s and 360s in the J$'$ and K$'$ filters, respectively. A summary of the observations is shown in Table \ref{Table1}. The observations show HD 38563C as an elongation due to the nebulae that the star is embedded. Although the FWHM of the observations were affected, the polarimetric measurements are not affected as these measurements were estimated in a large circular aperture assuming HD 38563C as a single source. In polarimetry, HD 38563C allows us to estimate the zero-angle calibration of the observations. The zero-angle calibration, $\Delta\theta$, was estimated as the difference of the measured P.A. of polarisation of our observations, $\theta_{\mbox{{\tiny J$'$}}}$ = 25\degree $\pm$ 5\degree~and $\theta_{\mbox{{\tiny K$'$}}}$ = 37\degree $\pm$ 2\degree, and the P.A. of polarisation, $\theta_{\mbox{{\tiny J}}}^{\mbox{{\tiny W}}}$ = 71\degree $\pm$ 1\degree~and $\theta_{\mbox{{\tiny K}}}^{\mbox{{\tiny W}}}$ = 78\degree $\pm$ 17\degree~provided by \citet{Whittet1992}. Thus, the zero-angle calibrations were estimated to be $\Delta\theta_{\mbox{{\tiny J$'$}}}$ = $\theta_{\mbox{{\tiny J}}}^{\mbox{{\tiny W}}}$ - $\theta_{\mbox{{\tiny J$'$}}}$ = 46\degree $\pm$ 5\degree~and $\Delta\theta_{\mbox{{\tiny K$'$}}}$ = $\theta_{\mbox{{\tiny K}}}^{\mbox{{\tiny W}}}$ - $\theta_{\mbox{{\tiny K$'$}}}$ = 41\degree $\pm$ 17\degree.  The instrumental polarisation was measured to be 0.05\% $\pm$ 0.03\% using several unpolarised standard stars from \citet{Whittet1992} during the observing run.


\begin{table}
\caption{Summary of Observations.}
\label{Table1}

\begin{tabular}{lcccc}
\hline

Object	&	Filter		&	Frame Time	&	\# ABA 	&	Total Time	\\
		&			&	(s)			&				&		(s)		\\
					
\hline

NGC 1068				&	J$'$		&		15				&		3						&		540		\\
						&	K$'$		&		10				&		10						&		1200		\\
HD 38563C			&	J$'$		&		10				&		2						&		240		\\								
						&	K$'$		&		10				&		3						&		360		\\

\hline
\end{tabular}
\end{table}



\section{Results}
\label{RES}

\begin{figure*}
\includegraphics[angle=0,trim=0cm 1cm 0cm 0cm,scale=.32]{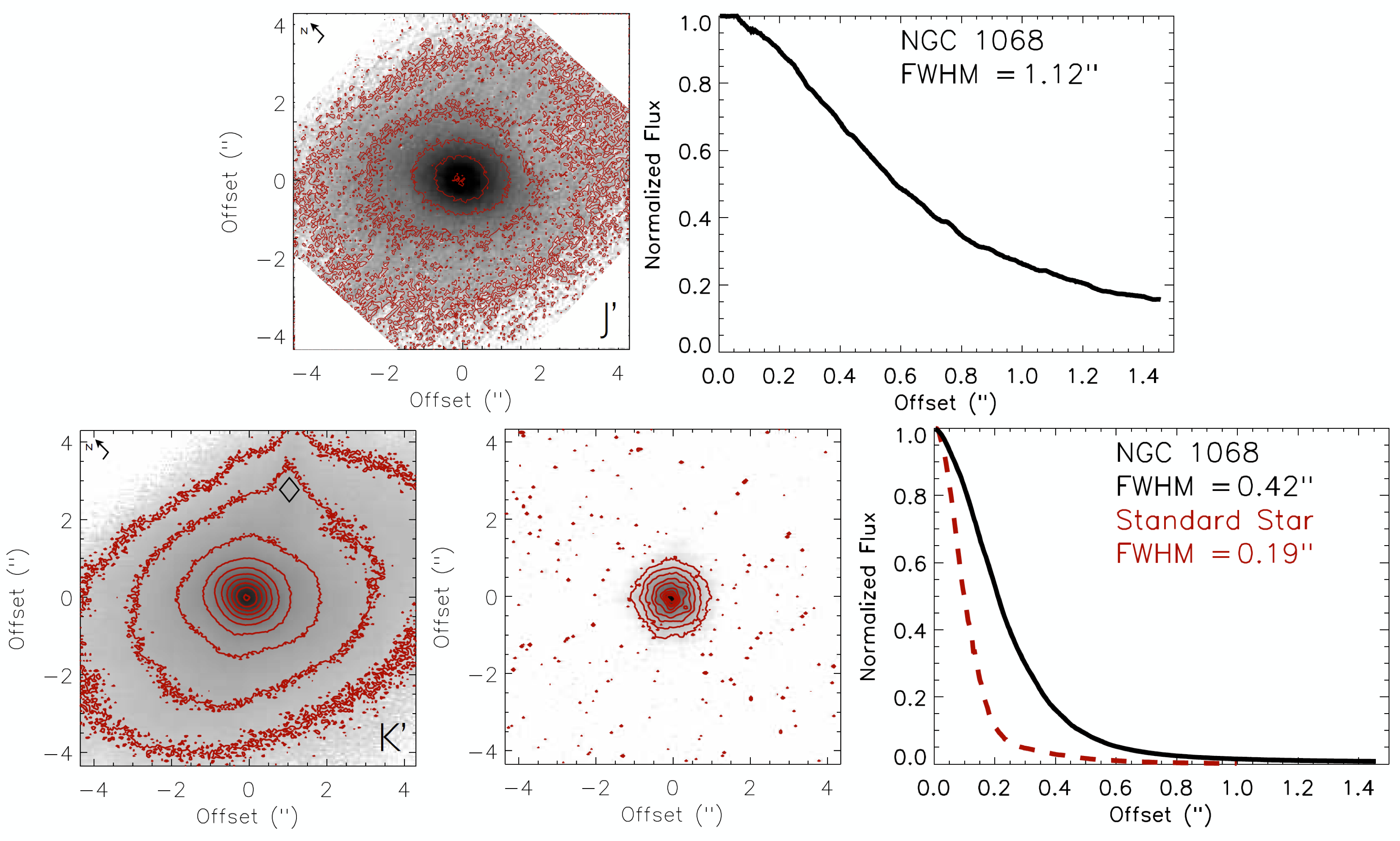}
\caption{The 8$\farcs$5 $\times$  8$\farcs$5 central region of NGC 1068 in the J$'$ (first row) and K$'$ (second row) total flux (grey scale) images. In both images, the lowest level contour is at 8$\sigma$, then contours are at the sigma-levels of 2$^{n}$, where $n$ = 4, 5, 6, etc. The J$'$ image was rotated for direct comparison with K$'$ image (Section \ref{OBS}). The ghost (black diamond) in the K$'$ is marked. The middle plot of the second row shows the unpolarised standard star, HD224467, at K$'$ with the same FOV and contour levels as the total flux images of NGC 1068. In all the total flux images, the pixel scale is 0.043$''$ pixel$^{-1}$. The radial profiles of NGC 1068 (black solid line) and the standard star (red dashed line), normalizes to the flux density value of the centroid of NGC 1068, are shown.
}
\label{fig1}
\end{figure*}

Figure \ref{fig1} shows the J$'$ (top-left) and K$'$ (bottom-left) total flux images of NGC 1068. The extension towards the West in the K$'$ total intensity image (Figure \ref{fig1}-black diamond) is a ghost which we attribute to faint, internal reflections within the optics. This faint ghost does not appear in polarised intensity. The radial profiles of the total flux images at J$'$ (top-right) and K$'$ (bottom-right) of NGC 1068 and the unpolarised standard star, HD224467, at K$'$ are shown. The FWHMs in the longest and shortest direction and associated position angle, using a 2D Gaussian profile, were estimated to be 1$\farcs$23 $\times$ 1$\farcs$01 at $P.A. = 56$\degree~E of N, and 0$\farcs$48 $\times$ 0$\farcs$38 at $P.A. = 50$\degree~E of N in the J$'$ and K$'$ filters, respectively. Note that the FWHMs in Figure \ref{fig1} are calculated using the radial profiles. The estimated FWHM of NGC 1068 differs from those ($\sim$0$\farcs$12 at 2.2 \um) previously reported in the literature using high-spatial resolution observations \citep[e.g.][]{Rouan1998,Rouan2004,G2006}. Although the FWHM of NGC 1068 was affected during the observations, only those observations with (1) photometric conditions; (2) locked AO in a complete ABA; and (3) good image quality, were used. Specifically, as the nucleus of NGC 1068 is faint, the AO system was forced to sample the optical image to lower frequencies compared to a bright star. At these lower frequencies, the AO system compensate less for seeing, exacerbated by the wind buffeting. To account for the impact of the weather condition and low AO camera frame rate in the measured flux, we estimated the variation of the counts  between the several ABA's to be 12\%. This result is included in the uncertainties of the AGN flux contribution in Section \ref{ANA_AGN}.


\begin{figure*}
\includegraphics[angle=0,trim=0cm 1cm 0cm 0cm,scale=.32]{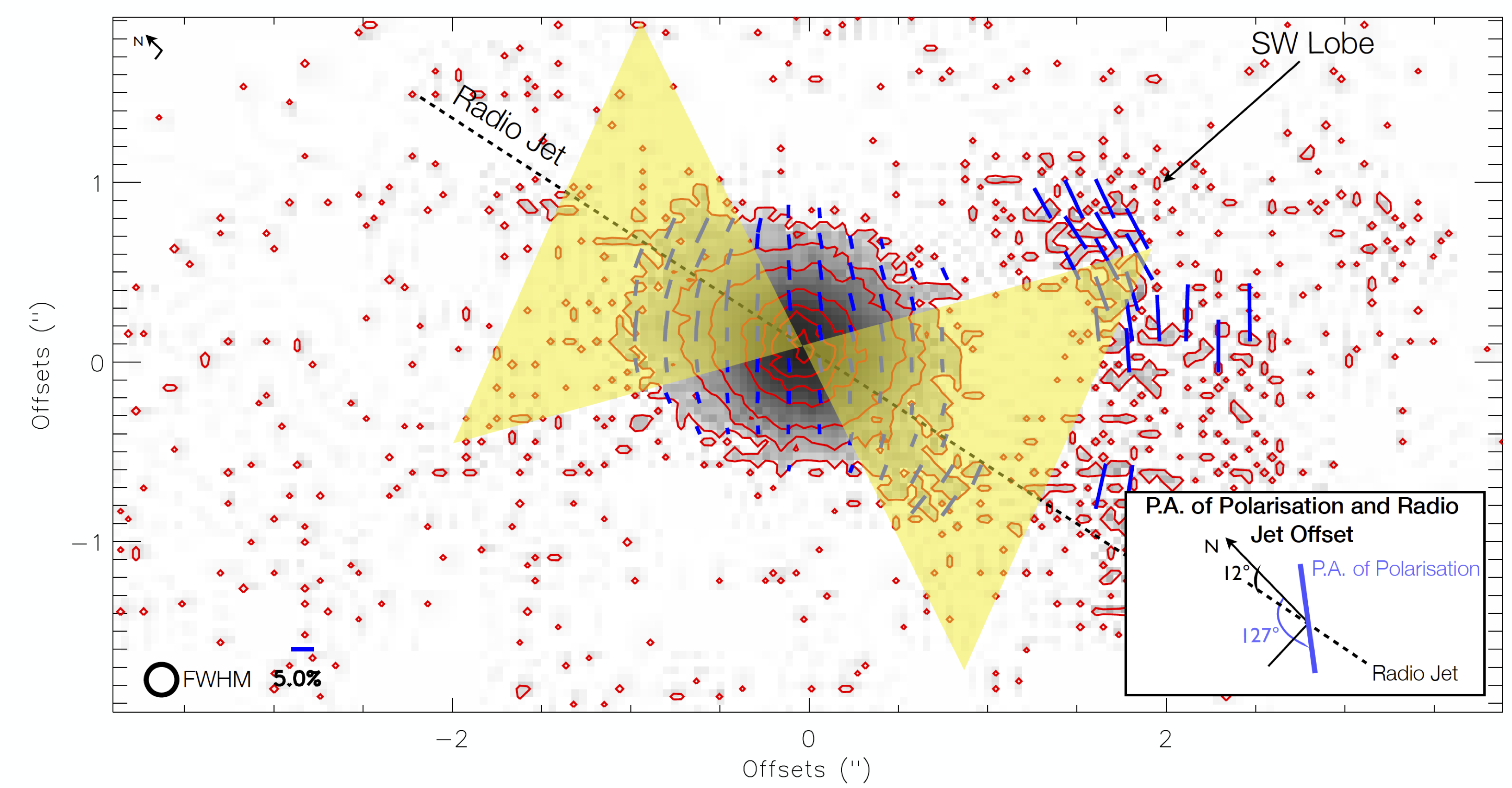}
\caption{The 8$''$ $\times$ 4$''$ polarised flux image (grey scale) at K$'$ with overlaid polarisation vectors. The contour levels are 1.5\%, 5\%, 10\%, 30\%, 50\%, 70\% and 90\% of the peak of the polarised flux image. The polarised flux image has a pixel scale of 0$\farcs$043, whilst the polarisation vectors with $P/\sigma_{p} > 3$ within the binned 4$\times$4 pixels box (0$\farcs$172) are shown. A vector of 5\% of polarisation and the FWHM of the observations are shown. The dashed line shows the orientation of the radio jet at P.A.$\sim$ 12\degree~by \citet{Gallimore1996}, and the yellow shadow area shows the {\it HST} [OIII] opening angle, 65\degree$\pm$20\degree~with P.A.$\sim$15\degree, by \citet{Evans1991}. The P.A. of polarisation and the radio jet axes (insert plot) with respect to the North is shown (Section \ref{RES}).}
\label{fig2}
\end{figure*}

Figure \ref{fig2} shows the polarised flux image and polarisation vectors through the K$'$ filter. The overlaid polarisation vectors are proportional in length to the degree of polarisation with their orientation showing the P.A. of polarisation. To improve the signal-to-noise ratio (S/N) and to obtain statistically independent polarisation measurements, the normalised Stokes parameters $q=Q/I$ and $u=U/I$ were binned from 4$\times$4 pixels (0$\farcs$172$\times$0$\farcs$172) to a single pixel. Then, only those polarisation vectors with $P/\sigma_{\mbox{\tiny p}} > 3$ are shown in Figure \ref{fig2}. $\sigma_{\mbox{\tiny p}}$ is the uncertainty on the degree of polarisation within the binned pixels, estimated as $\sigma_{\mbox{\tiny p}} = \sqrt{\sigma_{\mbox{\tiny u}}^2 + \sigma_{\mbox{\tiny q}}^2}$, where $\sigma_{\mbox{\tiny q}}$ and $\sigma_{\mbox{\tiny u}}$ are the uncertainties on the normalised Stokes parameters. In polarised flux, NGC 1068 shows a resolved nucleus with a FWHM estimated to be 0$\farcs$49 $\times$ 0$\farcs$45 at $P.A. = 30$\degree~E of N. The polarisation vectors and polarised flux image in the J$'$ filter are not shown because of the low S/N $\sim$ 40 at the peak pixel of the observations. 


To compare with previously published values (Table \ref{Table2}), we made measurements of the nuclear polarisation at J$'$ and K$'$ in several apertures. In all cases, polarimetric errors were estimated by variation of the measurements in subsets of the data. Our measured polarisation values in a 2$''$ aperture are in close agreement with previously published values (Table \ref{Table2}).  \citet{Simpson2002} measured a degree of polarisation of 6.0\% $\pm$ 1.2\%, with variations of $\pm$2-3\% in the degree of polarisation between the two observing runs of NGC 1068, however the reason of this difference is unknown. We subsequently conclude that the measured polarisation of \citet{Simpson2002} and ours, 4.8\% $\pm$ 0.1\%, are in marginal agreement within the uncertainties. We found an increase in the degree of polarisation with increasing wavelength, as expected from previous polarimetric studies \citep[e.g.][]{LKR78,Brindle1990,Young1995,Packham1997,Lumsden1999} with a roughly constant P.A. of polarisation in the NIR wavelength.

We also made measurements of the nuclear polarisation at J$'$ and K$'$ in a 0$\farcs$5 (30 pc) aperture. These measurements minimize contamination from extended (diffuse) dust emission, as well as obtain a statistically independent measurement with higher S/N than the single polarisation vectors (important in the J$'$ filter due to the low S/N of the observations). Specifically, its degree of polarisation was measured to be 2.9\% $\pm$ 1.3\% and 4.4\% $\pm$ 0.1\% with a P.A. of polarisation of 111\degree $\pm$ 14\degree~and 127\degree $\pm$ 2\degree~in a 0$\farcs$5 (30 pc) aperture in the J$'$ and K$'$ filters, respectively. The statistical significance of the degree of polarisation measurements is 2.4$\sigma$ and 36$\sigma$ in the J$'$ and K$'$ filters, respectively.


\begin{table}
\caption{Nuclear Polarimetry of NGC 1068 and Literature Values.}
\label{Table2}

\begin{tabular}{ccccc}
\hline

Aperture 	&	Filter	&	$P$ 	&	$P.A.$ 		&	Ref(s).			\\	
	($''$)	&			&	(\%)	&  (\degree)			&	\\

\hline

0.2						&		2.0 \um									&		6.0 $\pm$ 1.2			&	122 $\pm$ 15			& 		a  \\
							&		K$'$										&		4.8 $\pm$ 0.1			&	127 $\pm$ 2			&		b  	\\
0.5						&		J$'$										&		2.9 $\pm$ 1.3			&	111 $\pm$ 14			&		b 	\\		
						 	&		K$'$										&		4.4 $\pm$ 0.1			&	127 $\pm$ 2			&		b	\\
2.0						&		J											&		2.25 $\pm$ 0.25		&	106.8 $\pm$ 2.0		&		c 		\\
							&		J											&		1.71 $\pm$ 0.22		&	106.45 $\pm$ 1.70	&		d 		\\
							&		J$'$										&		2.4 $\pm$ 1.2			&		106 $\pm$ 17	&		b		\\
							&		K$_{\mbox{{\tiny n}}}$				&		4.57 $\pm$ 0.50		&	120.2 $\pm$ 2.0		&		c		\\
							&		K$_{\mbox{{\tiny n}}}$				&		4.11 $\pm$ 0.46		&	120.55 $\pm$ 2.38		&		d		\\	
							&		K$'$										&		4.2 $\pm$ 0.3			&		119 $\pm$ 5		& 		b		\\
\hline
\end{tabular}\\
References: (a) \citet{Simpson2002}; (b) This work; (c) \citet{Lumsden1999}; (d) \citet{Packham1997}
\end{table}


The polarised flux image at K$'$ shows a centrosymmetric polarisation pattern (SouthWest Lobe in Figure \ref{fig2}) at $\sim$1$\farcs$5 South from the center of the galaxy. The degree of polarisation is 15\% $\pm$ 2\% within the binned 4$\times$4 pixels (0$\farcs$172$\times$0$\farcs$172). This polarisation pattern is the signature of a central source whose radiation is scattered by dust/electrons. This polarised structure and measured polarisation are in excellent agreement with (1) the 2.0 \um~polarimetric observations using {\it HST}/NICMOS by \citet{Simpson2002}, and (2) the ionisation cones, with opening angle of 65\degree$\pm$20\degree~and P.A.$\sim$15\degree, observed using {\it HST}  [OIII] observations by \citet{Evans1991}. We found a $\sim$49\degree~misalignment between the measured P.A. of polarisation at K$'$ and the radio jet axis (Figure \ref{fig2}, insert plot). This misalignment is similar to the $\sim$45\degree~misalignment between the compact (1.35 $\times$ 0.45 pc) hot ($\sim$800 K) dust component (i.e. obscuring torus) observed by \citet{Raban2009} and the radio jet axis. A further interpretation of this misalignment in terms of the MHD outflow wind model in AGN is given in Section \ref{DIS_B}.



\section{Analysis}
\label{ANA}


\subsection{AGN contribution to the total flux density}
\label{ANA_AGN}

The emission from the unresolved core of NGC 1068 at K$'$ can arise from two components: (1) diffuse stellar emission in the nuclear bulge, and (2) AGN emission. We interpret the AGN emission as a point-source emission component, while the diffuse stellar emission is extended. Photometric cuts through the nucleus in the J$'$ filter show negligible evidence of a nuclear point-source, AGN emission, due to extinction towards the core and/or contamination by the diffuse stellar emission (Figure \ref{fig1}, top-right). The J$'$ filter profile is assumed to be representative of the diffuse stellar emission in the nuclear bulge. At K$'$, both the AGN emission and diffuse stellar emission are detectable. To estimate the relative contributions from both emission components in the K$'$ filter, two different methodologies were followed. 

In the first method (method 1), we followed similar analysis to that of \citet{Turner1992,Packham1996} and \citet{LR2013}. We took photometric profiles along the major axis of the diffuse extended stellar emission in the J$'$ and K$'$ filters, which is shown to have a constant extinction of  A$_{\mbox{{\tiny V}}}$ = 0.7 $\pm$ 0.7 \citep{Young1995}. These cuts allow us to characterize the host galaxy contribution. The  contribution of the AGN emission is assumed to be the point-spread function (PSF) of the observations (i.e. the unpolarised standard star, HD224467). To fit the K$'$-profile, the emission was modeled as the sum of a scaled-PSF and the diffuse stellar emission from the J$'$-profile. Using this method the best estimate of AGN contribution to the total flux is 60\% $\pm$ 12\% in a 0$\farcs$5 (30 pc) aperture.  The uncertainty was estimated as the 12\% uncertainty due to weather conditions and low AO camera frame rate (Section \ref{RES}).

In the second method (method 2), the PSF from method 1 was scaled to the flux density of the K$'$ emission to subtract the AGN emission. This procedure produces a $``$flat-top$"$ profile over the unresolved core of NGC 1068 \citep[for examples of the use of this technique, see][]{Radomski2002,Radomski2003,RA11,LR2013}. Using this method, the contribution of the AGN emission to the total flux is 50\% $\pm$ 19\% in a 0$\farcs$5 (30 pc) aperture. The uncertainty was estimated as the addition in quadrature of the uncertainty of the $``$flat-top$"$ method applied, 15\%, and the 12\% uncertainty due to weather conditions and low AO camera frame rate (Section \ref{RES}).

From both methods presented above, the average of the AGN emission to the total flux is estimated to be 55\% $\pm$ 16\%, whereas the contribution of the diffuse stellar emission to the total flux is 45\% $\pm$ 16\%, in a 0$\farcs$5 (30 pc) aperture in the K$'$ filter. The formal total uncertainties are 16\% through the average of both methods. Systemic errors of methodologies could increase these estimates, but are difficult to quantify. For NGC 1068, \citet{Weinberger1999} estimated the AGN contribution to be 49\% $\pm$ 8\% of the total nuclear flux in a 0$\farcs$48 aperture in the K filter using the near infrared camera (NIRC) on the 10-m Keck I telescope. \citet{Weigelt2004} estimated a total flux contribution of 46\% $\pm$ 5\% at K$'$ in a 18 $\times$ 39 mas using NIR speckle interferometry on the 6-m Special Astrophysics Observatory (SAO).


\subsection{The nuclear intrinsic polarisation}
\label{ANA_INT_POL}

The diffuse stellar emission in the nuclear regions of NGC 1068 significantly dilutes the observed polarisation from the AGN. If only the dilution from the diffuse stellar emission is subtracted from the observed polarisation, the estimated intrinsic polarisation is independent of the dominant polarisation mechanism in the nuclear regions of NGC 1068. Thus, the dominant polarisation mechanism in the K$'$ filter is needed to estimate the intrinsic degree of polarisation of the AGN. The intrinsic degree of polarisation arising from the dominant mechanism of polarisation in the nuclear regions of NGC 1068 is estimated as follows. The measured polarisation at K$'$, $P^{\mbox{{\tiny obs}}}_{\mbox{{\tiny K$'$}}}$ = 4.4\% $\pm$ 0.1\%, in a 0$\farcs$5 (30 pc) aperture is corrected by accounting for (1) the measured degree of polarisation through dichroic absorption of starlight in an off-nuclear region ($P_{\mbox{\tiny K$'$}}^{\mbox{\tiny off}}$), (2) the diffuse stellar emission relative to the AGN emission in the nuclear aperture ($R_{\mbox{\tiny K$'$}}$), and (3) the dominant polarisation mechanism to the polarised flux of the AGN in the nuclear aperture ($F^{\mbox{{\tiny dic}}}_{\mbox{{\tiny K$'$}}}$).

The degree of polarisation at K$'$ through dichroic absorption of starlight in the off-nuclear regions of NGC 1068 was measured to be $P^{\mbox{{\tiny off}}}_{\mbox{{\tiny K$'$}}}$ = 0.5\% $\pm$ 0.3\%. The off-nuclear polarisation was estimated as the average of several polarisation measurements using a 2$''$ aperture. We assume that the off-nuclear polarisation has the same P.A. of polarisation as the AGN polarisation. Any value of the P.A. of polarisation will produce negligible variations in the estimation of the intrinsic polarisation, given the low measured polarisation. Using the diffuse stellar emission, $I^{\mbox{{\tiny ste}}}_{\mbox{{\tiny K$'$}}} =$ 45\% $\pm$ 16\%, relative to the AGN emission, $I^{\mbox{{\tiny AGN}}}_{\mbox{{\tiny K$'$}}}$ = 55\% $\pm$ 16\% estimated in Section \ref{ANA_AGN}, we define the ratio of both emission components as $R_{\mbox{{\tiny K$'$}}} = I^{\mbox{{\tiny ste}}}_{\mbox{{\tiny K$'$}}} / I^{\mbox{{\tiny AGN}}}_{\mbox{{\tiny K$'$}}}$ = 0.8$\pm$0.6.  

As noted in the introduction, previous studies \citep{Young1995,Packham1997,Lumsden1999,Simpson2002,Watanabe2003} have shown that the NIR polarisation in the nuclear regions of NGC 1068 arises from the passage of light through aligned dust grains in the torus. As we only have polarisation measurements in two NIR filters, we cannot estimate the relative contribution of the dichroic absorption to the polarised flux at K$'$. We therefore resort to previously published results. \citet{Young1995} found that the dichroic absorption mechanism accounts for $\sim$90\% of the polarised flux, with a combination of electron and dust scattering from the ionisation cones accounting for $\sim$10\% of the polarised flux at 2.0 \um~in a 3$\farcs$08 $\times$ 3$\farcs$00 (185 $\times$ 180 pc) slit aperture \citep[figure 6 from][]{Young1995}. \citet{Watanabe2003} found that the dichroic absorption mechanism can account for $\sim$100\% of the polarised flux at 2.2 \um~in a 0$\farcs$88 $\times$ 2$\farcs$00 (53 $\times$ 120 pc) slit aperture (figure 1 there). These slit sizes are larger than the 0$\farcs$5 (30 pc) aperture in our study. The difference in aperture sizes can affect the polarisation measurements due to scattered emission from the extended biconical structures around the core of NGC 1068 \citep{Packham1997,Lumsden1999,Simpson2002}, and the diffuse stellar emission in the host galaxy. \citet{Simpson2002} showed that the polarised flux from the core of NGC 1068 completely dominates within an aperture $<$1$''$, where contributions from off-nuclear scattered light would be barely detectable. Other polarisation components (i.e. electron scattering from the inner regions of the ionisation cones and/or dust scattering by large grains in the torus) can contribute, at some level, to the polarised flux at K$'$ within our 0$\farcs$5 (30 pc) aperture, however the contribution of these components are highly dependent of the geometry of the system, implying a large amount of assumptions that will introduce high uncertainties in the estimation of the intrinsic polarisation. Based on these studies, these polarisation mechanisms cannot be ruled out, but the contribution to the polarised flux of these components have not been quantified yet and require more detailed polarisation models. We here assume the upper limit of the dichroic absorption to be 100\% of the polarised flux, in agreement with the polarisation model of \citet{Watanabe2003}, i.e. $F^{\mbox{{\tiny dic}}}_{\mbox{{\tiny K$'$}}}$ = 1. The intrinsic polarisation at K$'$ arising from dichroic absorption, $P^{\mbox{{\tiny int}}}_{\mbox{{\tiny K$'$}}}$, is

\begin{equation}
P^{\mbox{{\tiny int}}}_{\mbox{{\tiny K$'$}}} = (P^{\mbox{{\tiny obs}}}_{\mbox{{\tiny K$'$}}} - P^{\mbox{{\tiny off}}}_{\mbox{{\tiny K$'$}}}) \times (1 + R_{\mbox{{\tiny K$'$}}}) \times F^{\mbox{{\tiny dic}}}_{\mbox{{\tiny K$'$}}}
\label{eq_Pint}
\end{equation}
\noindent
and is estimated to be $P^{\mbox{{\tiny int}}}_{\mbox{{\tiny K$'$}}}$ = 7.0\% $\pm$ 2.2\% in a 0$\farcs$5 (30 pc) aperture. 


\section{Magnetic Field strength in the Torus}
\label{MAG}

The magnetic field strength in the NIR emitting regions of the torus of NGC 1068 is estimated following similar analysis to that of \citet{LR2013}. Specifically, the magnetic field strength is estimated through three different methods: (1) paramagnetic alignment, (2) thermal and magnetic relaxation time equipartition, and (3) Chandrasekhar-Fermi method. 


\subsection{Physical conditions in the torus of NGC 1068}
\label{MAG_cond}

NIR reverberation mapping of several AGNs has shown that the outer radius of the broad line region (BLR) approximately corresponds to the inner radius of the dusty torus \citep{Suganuma2006,K2014}. The gas temperature reaches a value of $\sim$10$^{4}$ K in the BLR \citep{Netzer1987}.  \citet{KK2001} suggested that a warm absorber gas in the inner edge of the torus can reach temperatures in the range of 10$^{4}-10^{6}$ K. Based on these previous studies, we adopt the lower-limit of the gas temperature to be  $T_{\mbox{\tiny gas}}$ = 10$^{4}$ K.
We here adopt the \textsc{Clumpy} torus model\footnote{For details of the \textsc{Clumpy} torus models see www.clumpy.org} by \citet{Nenkova2002,Nenkova2008a,Nenkova2008b}. This model depicts the torus as a distribution of optically thick and dusty clouds surrounding the central engine, instead of homogeneously filling the torus volume. The NIR-emitting region in the torus is identified with clumps directly illuminated by the central engine, with dust grain temperatures below the sublimation temperature in the range of  800$-$1500 K. The dust grain sizes are assumed to be in the range of 0.005$-$0.25 \um. Based on the \textsc{Clumpy} torus description and the further fitting to the nuclear IR SED of NGC 1068 by \citet{AH2011} using \textsc{BayesClumpy} \citep{AA2009}, the torus of NGC 1068 can be described with a radial thickness of $Y =$ 6$^{+2}_{-1}$, angular width of $\sigma =$ 26$^{+6}_{-4}$\degree, number of clouds along the equatorial $N_{\mbox{\tiny 0}} =$ 14$^{+1}_{-3}$, radial density profile with an index of $q =$ 2.2$^{+0.4}_{-0.3}$, viewing angle of $i =$ 88$^{+2}_{-3}$\degree~(almost edge-on), and the optical depth per single cloud of $\tau_{\mbox{\tiny V}} =$ 49$^{+4}_{-3}$. The distance of the inner wall of the torus from the black hole, $r_{\mbox{\tiny sub}}$, is assumed to be the sublimation distance given by $r_{\mbox{\tiny sub}} = 1.3 (L_{\mbox{\tiny bol}}/10^{46} \mbox{erg s$^{-1}$})^{1/2} (T_{\mbox{\tiny gr}}/1500 \mbox{K})^{-2.8}$ pc \citep{Barvainis1987}. We took the bolometric luminosity of NGC 1068 to be L$_{\tiny{bol}}$ = $9.55~\times~10^{44}$ erg s$^{-1}$ by \citet{WU2002}, and we estimated $r_{\mbox{\tiny sub}} =$ 0.4 pc for dust grains at 1500 K. The torus height is estimated as $H = r_{\mbox{\tiny out}} \sin{\sigma}$ = 1.1$^{+0.7}_{-0.3}$ pc, with $r_{\mbox \tiny out} = Yr_{\mbox{\tiny sub}}$ = 2-3.2 pc. The half-opening angle of the torus is estimated as $\Theta = 90\mbox{\degree} - \sigma = 64^{+4}_{-6}$\degree, which is defined with respect to the projected symmetry axis of the \textsc{Clumpy} model. The number density in the individual clumps of NGC 1068 were calculated assuming the output parameters of the \textsc{Clumpy} torus models by \citet{AH2011}: (1) a torus radius of $r_{\mbox{\tiny out}} =$ 2.0-3.2 pc, (2) the number of clouds along the equatorial direction of $N_{\mbox{\tiny 0}} =$ 14$^{+1}_{-3}$, and (3) the optical depth per cloud of $\tau_{\mbox{\tiny v}} =$ 49$^{+4}_{-3}$, converted to be $A_{\mbox{\tiny v}} = 1.086\tau_{\mbox{\tiny v}} = $ 53$^{+4}_{-3}$ mag. Using the standard Galactic ratio $A_{\mbox{\tiny v}}/N_{\mbox{\tiny H}} =$ 5.23 $\times$ 10$^{-22}$ mag cm$^{-22}$ \citep{BSD78}, the column density of the individual clouds is $N_{\mbox{\tiny H}} =$ 1.01$^{+0.08}_{-0.06} \times$ 10$^{23}$ cm$^{-2}$. The number density in the individual clouds in the torus are estimated to be n = $N_{\mbox{\tiny H}} / (r_{\mbox{\tiny out}}/N_{\mbox{\tiny 0}} ) =$ 2.30$^{+0.36}_{-0.61}$ $\times$ 10$^{5}$ cm$^{-3}$, which compare very well with the number density in the range of 10$^{4}$ $-$ 10$^{5}$ cm$^{-3}$ for molecular clouds in Orion A, M17 and Cepheus A \citep[e.g.][]{Goldsmith1999}. A summary of the physical parameters of the torus is shown in Table \ref{Table3}.


\begin{table*}
\caption{Physical Parameters of the Torus of NGC 1068 }
\label{Table3}

\begin{tabular}{lcr}
\hline

Description	&	Parameter	&	Value	\\							
\hline

Radial extent of the torus										&	$Y$			&	6$^{+2}_{-1}$	\\
Width of the angular distribution of clouds					&	$\sigma$		&	26$^{+6}_{-4}$\degree	\\
Number of clouds along the radial equatorial direction	&	$N_{\tiny 0}$	&	14$^{+1}_{-3}$	\\
Power-law index of the radial density profile				&	$q$			&	2.2$^{+0.4}_{-0.3}$	\\
Inclination of the torus											&	$i$				&	88$^{+2}_{-3}$\degree	\\
Optical depth per single cloud									&	$\tau_{\mbox{\tiny V}}$	&	49$^{+4}_{-3}$	\\
Sublimation radius					& $r_{\mbox{\tiny sub}}$		&  0.4 pc	\\	
Torus height							& $H$								& 1.1$^{+0.7}_{-0.3}$ pc	\\
Half-opening angle					& $\theta$						&  64$^{+4}_{-6}$\degree	\\	
Gas temperature    					& $T_{\mbox{\tiny gas}}$		&  $10^{4}$  K  \\
Grain temperature					&  $T_{\mbox{\tiny gr}}$		&  800 $-$ 1500 K      \\
Grain size		 					& $a$								&   0.005 $-$ 0.25 \um                   \\
Number density in the clumps 	&  $n$								&  2.30$^{+0.36}_{-0.61}$ $\times$ 10$^{5}$ cm$^{-3}$       \\

\hline
\end{tabular}
\end{table*}



\subsection{Method 1: Paramagnetic alignment}
\label{MAG_M1}

We concluded that the NIR polarisation arises from the passage of radiation through aligned dust grains in the clumps of the torus of NGC 1068 (Section \ref{ANA_INT_POL}). Non-spherical spinning dust grains can become aligned in the presence of a magnetic field, a process termed paramagnetic alignment \citep{DG1951}. If the dust grains are aligned by this mechanism, then some degree of polarisation through dichroic absorption and/or emission can be measured. This measurement can be used to estimate the strength and geometry of the magnetic field. In case of the geometry, the long axes of the dust grains are aligned perpendicular to the direction of the magnetic field. A P.A. of polarisation parallel to the magnetic field is expected for dichroic absorption. In case of the magnetic field strength, if paramagnetic alignment is the dominant mechanism, then the dust grain alignment efficiency is given by the ratio of the degree of polarisation, $P$(\%), and the extinction, $A_{\mbox{\tiny V}}$, and is related to the magnetic field strength, $B$, as shown by \citet{JS1967}, and adapted by \citet{Vrba1981}, and \citet{LR2013}: 

\begin{align}
P(\%)/A_{\mbox{\tiny V}} = \frac{67 \chi'' B^2}   {75 a \omega n}  \left( \frac{2 \pi} {m_{\mbox{\tiny H}} k T_{\mbox{\tiny gas}}} \right)^{1/2}  (\gamma - 1) \nonumber \\
\times \left( 1 - \frac{T_{\mbox{\tiny gr}}}{T_{\mbox{\tiny gas}}}  \right)
\label{P/Av}
\end{align}
\noindent
$\chi''$ is the imaginary part of the complex electric susceptibility, a measure of the attenuation of the wave caused by both absorption and scattering, $B$ is the magnetic field strength, $\gamma$ is the ratio of inertia momentum around the two major axes of the dust grains, $T_{\mbox{\tiny gr}}$  is the dust grain temperature, $a$ is the dust grain size,  $\omega$ is the orbital frequency of the dust grains, $n$ is the number density in the cloud, $m_{\mbox{\tiny H}}$ is the mass of a hydrogen atom, $k$ is the Boltzmann constant, and $T_{\mbox{\tiny gas}}$ is the gas temperature.

\citet{DG1951} showed that the lower bound of the ratio $\chi''$/$\omega$ for most interstellar grains is:

\begin{equation}
\frac{\chi''}{\omega} = 2.5 \times 10^{-12}~T_{\mbox{\tiny gr}}^{-1}
\end{equation}

The ratio of the moments of inertia of the dust grains, $\gamma$, is defined as:
\begin{equation}
\gamma = \frac{1}{2} \left[ \left( \frac{b}{a} \right)^{2} + 1  \right]
\end{equation}
\noindent
where $b/a$ is the dust grain axial ratio.  A typical value of $b/a$ for interstellar dust grains is $\sim$ 0.2 \citep{AP73,KM95}. It should be noted that the real dust composition in AGN might be more complex that the typical interstellar dust grains, as several authors suggest \citep[e.g.][]{M2001,MMO2001,H2015}

The magnetic field strength is estimated using the physical conditions in Table \ref{Table3}, the intrinsic polarisation at K$'$ arising from dichroic absorption of $P^{\mbox{{\tiny int}}}_{\mbox{{\tiny K$'$}}}$ = 7.0\% $\pm$ 2.2\% (Section \ref{ANA_INT_POL}), and the extinction to the central engine $A_{\mbox{{\tiny V}}}$ $=$ 36 mag \citep{Packham1997} converted\footnote{The conversion of visual extinction, $A_{\mbox{{\tiny V}}}$, to extinction at K, $A_{\mbox{{\tiny K}}}$, is $A_{\mbox{{\tiny K}}}$ = 0.112$A_{\mbox{{\tiny V}}}$ \citep{Jones1989}.} to $A_{\mbox{{\tiny K$'$}}}$ $=$ 4.03 mag. The extinction represents the closest measurement at our observations, and it has been previously used by \citet{Watanabe2003}. \citet{Imanishi1997} estimated a visual extinction in the range of 17$-$30 mag using the 3.4 \um~absorption feature of the NIR (2.9$-$4.1 \um) spectroscopic observations on the 1.5-m telescope on Mount Lemmon, Arizona, USA. These measured visual extinctions are lower than the whole visual extinction of the clouds, $A_{\mbox{\tiny v}} = 53^{+4}_{-3}$ mag. (Section \ref{MAG_cond}), estimated using the \textsc{Clumpy} model to the IR SED of NGC 1068, indicating that we are possibly seeing the radiation from the clouds and/or central engine passing through a section of the clouds (Section \ref{DIS_B}). Based on the range of the physical parameters shown in Table \ref{Table3}, the magnetic field strength is estimated to be in the range of 4$-$82 mG for the NIR emitting regions in the torus of NGC 1068. The magnetic field strength in the maser emission of NGC 1068  was estimated to be 71 mG and 3 mG for standard accretion disc and advection-dominated accretion flow (ADAF) model, respectively \citep{Gnedin14}. Several studies \citep[e.g.][]{H1998,K1999,Modjaz2005} have shown that the magnetic field strength is in the range of 2$-$300 mG at scales of 0.1$-$1 pc from the central engine of AGN. Although these measurements were done for different objects, our estimated magnetic field strength in the torus of NGC 1068 is in agreement with the typical magnetic field strengths in the surrounding areas of the accretion disc in AGNs. 

Several assumptions were made in the estimation of the magnetic field strength in the torus of NGC 1068. Here, we consider each of these assumptions in detail. The physical conditions in molecular clouds make the alignment of dust grains less responsive to the magnetic field \citep[i.e.][]{Lazarian1995,Gerakines1995}. However, the dust grains in the dusty torus of AGN certainly experience much more turbulence and extreme physical conditions than in molecular clouds, and the dust alignment is more sensitive to the magnetic field strength (Section \ref{MAG_M2}, \ref{DIS_B}).

We assumed a homogeneous magnetic field, where any inhomogeneities of the magnetic field in the torus are ignored. This assumption has some implications for the ratio $P$(\%)/$A_{\mbox{\tiny V}}$. If a homogeneous magnetic field is responsible for dust grains' alignment in the torus, then all the dust grains will be aligned along the same direction of the magnetic field. In this case, the alignment efficiency, $P$(\%)/$A_{\mbox{\tiny V}}$, is maximal. The degree of polarisation would decrease when inhomogeneities of the magnetic field are present (Section \ref{DIS_B}). This method shows a strong dependence on dust grain sizes. The grain size distribution in the clumps is poorly constrained, which makes it difficult to quantify the uncertainties introduced in the estimations of the magnetic field strength.


\subsection{Method 2: Thermal and magnetic relaxation time equipartition}
\label{MAG_M2}

The approach followed in Method 1 (Section \ref{MAG_M1}) is valid only when the ratio of thermal to magnetic pressure is smaller than unity \citep{Gerakines1995}, meaning that the magnetic field is dominant. Under this condition, the magnetic field is strong enough to align the dust grains faster than the rotational kinetic energy. The magnetic relaxation time is required to be shorter than the thermal relaxation time, i.e. $t_{\mbox{\tiny m}}$ $<$  $t_{\mbox{\tiny th}}$. To verify this condition, we calculate the required lower limit magnetic field strength given by \citep[section 7.2 from][]{LR2013}:

\begin{equation}
B^2 > 2.4 \times 10^{11} a n m_{\mbox{\tiny H}} T_{\mbox{\tiny gr}} \left( \frac{8kT_{\mbox{\tiny gas}}}{\pi m_{\mbox{\tiny H}}} \right)^{1/2}
\label{B}
\end{equation}

The magnetic field strength is estimated to be $>$6 mG for the physical conditions shown in Table \ref{Table3}. Under the condition that the magnetic relaxation time is shorter than the thermal relaxation time, the gas and dust are decoupled. This result has two implications: (1) it is only valid in low-density regions of the clouds, and (2) the ratio $P$(\%)/$A_{\mbox{\tiny v}}$ is dependent on the magnetic field strength. Thus, we conclude that the detected NIR polarisation arises from the passage of NIR radiation through the low-density regions of the clumps in the torus. Otherwise, the NIR radiation would be completely extinguished (Section \ref{DIS_B}). This scheme satisfies condition (1) and the estimation of the magnetic field strength in method 1 (Section \ref{MAG_M1}) is allowed.


\subsection{Method 3: Chandrasekhar-Fermi method}
\label{MAG_M3}

The Chandrasekhar-Fermi method \citep[][hereafter CF method]{CF1953} gives us an alternative estimation of the magnetic field strength. Specifically, the CF method relates the magnetic field strength with the dispersion in polarisation angles, $\alpha$, of the constant component of the magnetic field, and the velocity dispersion of the gas. We use the adapted version of \citet{OSG2001}, where they introduced the factor of 0.5 in the CF method to compensate for averaging along the LOS. 

\begin{equation}
B = 0.5 \left( \frac{4}{3}\pi\rho \right)^{1/2}\frac{\sigma_{\mbox{v}}}{\alpha}~~[\mu \mbox{G}]
\label{CF_eq}
\end{equation}

\noindent
where $\rho$ is the volume mass density in g cm$^{-3}$, $\sigma_{\mbox{v}}$ is the velocity dispersion in cm s$^{-1}$, and $\alpha$ is the dispersion of polarisation angles in radians. The volume mass density was calculated using the number density in Table \ref{Table3} multiplied by the weight of molecular hydrogen. \citet{Greenhill1996} estimated velocity dispersions up to 100 km s$^{-1}$ of the VLBI H$_{2}$O masers at scales of 0.65 pc from the central engine of NGC 1068 using VLBI. We note that these results are higher than the typical velocity dispersion of 10 km s$^{-1}$ used in previous MHD outflow wind models \citep[e.g.][]{ES2006}, which can yield higher estimations of the magnetic field strength.


\begin{figure}
\includegraphics[angle=0,trim=1cm 1cm 0cm 0cm,scale=.28]{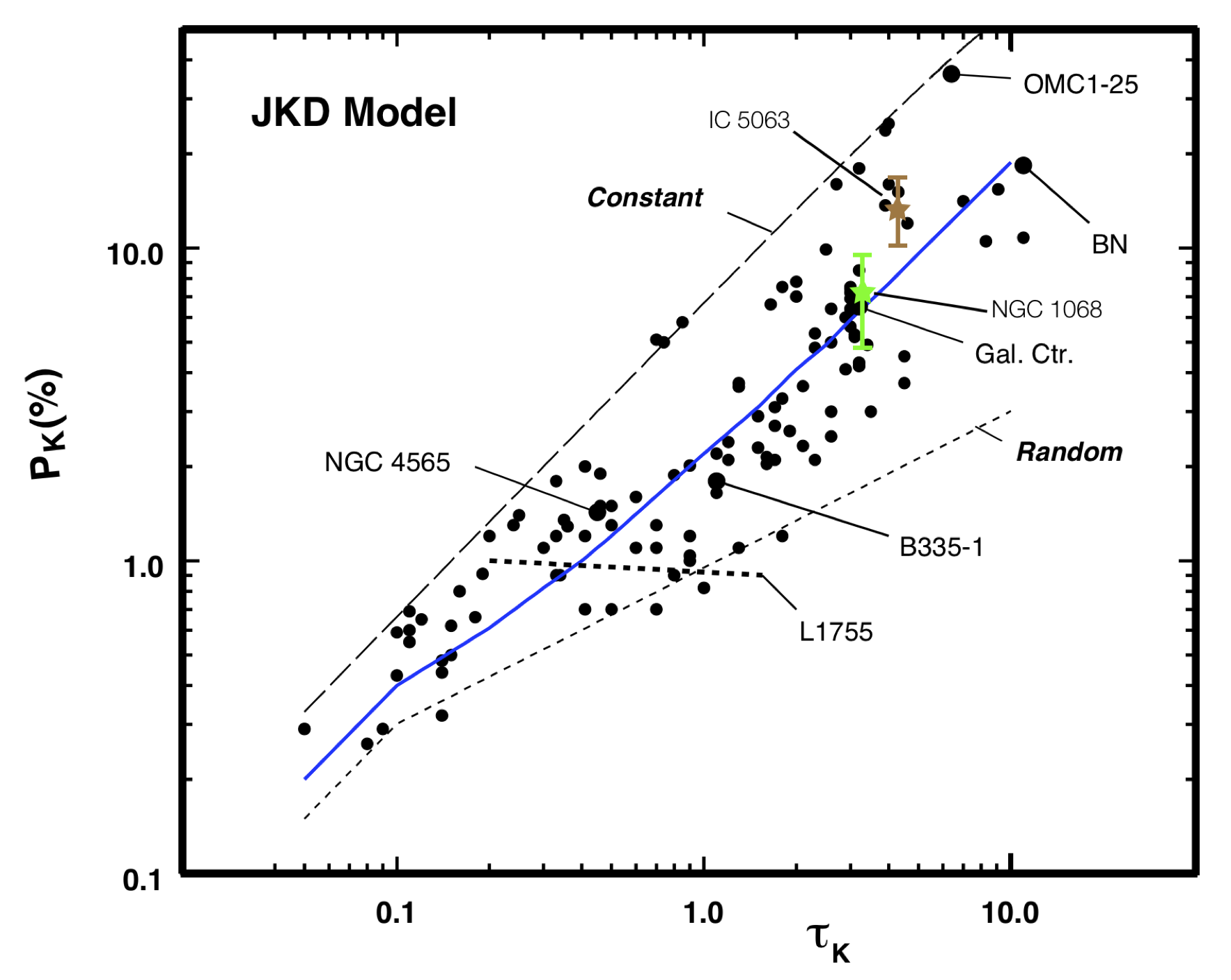}
\caption{
Degree of polarisation at K versus optical depth. Plotted is the data compilation from \citet{Jones1989} (black dots) with some specific objects labeled. The upper (long dash) line is the JKD model result assuming only a constant component to the magnetic field $(P = \tanh( \tau_P))$. The bottom (short dash) line corresponds to a purely random magnetic field with a decorrelation length of $\tau_K = 0.1$ $(P \propto \tau^{1/2})$. The solid blue line results from a 50/50 mix of constant and random components and fits the overall trend in the data well. The data point for IC 5063 is plotted as a brown star \citep{LR2013}. The estimated intrinsic polarisation for NGC 1068 from this work, $P_{K'}^{{\mathop{\rm int}} } = 7.0 \pm 2.2 \%$ (Section \ref{ANA_INT_POL}) with $A_V=36$ from \citep{Packham1997}, is plotted as a green star.
}
\label{fig3}
\end{figure}


As the AGN in NGC 1068 is unresolved, the polarisation vectors are average values of the LOS to the nucleus. If we average the LOS across the observed core (Figure \ref{fig2}), it will provide an insignificant dispersion in P.A. This measurement will yield unrealistically large magnetic field strengths with the CF method. Hence, the dispersion of P.A., $\alpha$, cannot be measured directly from the observed dispersion of the polarisation vectors across the core of NGC 1068 at K$'$. We need another approach to estimate $\alpha$; we used the model by \citet[][hereafter JKD]{JKD1992}. This model relates the degree of polarisation at K with the level of turbulence in the interstellar medium and the magnetic field. The magnetic field is assumed to have a constant and a random component. For NGC 1068, using the intrinsic polarisation arising from dichroic absorption at K$'$, $P^{\mbox{{\tiny int}}}_{\mbox{{\tiny K$'$}}} $= 7.0\% $\pm$ 2.2\% (Section \ref{ANA_INT_POL}), and the extinction to the central engine, $A_{\mbox{{\tiny v}}}$ $=$ 36 mag \citep{Packham1997}, converted\footnote{The conversion of visual extinction, $A_{\mbox{{\tiny K}}}$, to optical depth at K, $\tau_{\mbox{{\tiny K}}}$ = 0.09$A_{\mbox{{\tiny v}}}$ \citep{Jones1989}} to be $\tau_{\mbox{{\tiny K}}}$ = 3.24, we found that our data in the JKD model is close to equal contribution of the constant and random components of the magnetic field (Figure \ref{fig3}). For our data point in Figure \ref{fig3}, if we assume that the constant component of the magnetic field is in the plane of the sky, the dispersion of polarisation angles is estimated to be $\alpha$ = 7.0\degree~(0.1221 radians) using figure 9 by \citet{JKD1992}. We substituted the above numerical values in Equation (\ref{CF_eq}), and estimated a lower limit of the magnetic field strength in the plane of the sky in the range of 52$^{+4}_{-8}$ mG, depending on the conditions in the torus.

We assumed that the constant component of the magnetic field strength is in the plane of the sky. If the constant component of the magnetic field is away from the plane of the sky, then the magnetic field strength will be underestimated. For example, if the magnetic field is pointing along our LOS, zero polarisation will be measured. Since NGC1068 is classified as a Type 2 AGN, the torus axis is in the plane of the sky. This geometry favors the assumption of the constant component of the magnetic field to be in the plane of the sky. We note that the all-sky \textit{Planck} polarisation data \citep{Planck2014} shows some objects with polarisation measurements above the theoretical polarisation maximum in the JKD model. However, it is difficult to compare the \textit{Planck} results with ours, given that \textit{Planck}'s beam of 5$'$ does not resolve AGNs.

A summary of the estimations of the magnetic field strengths through the different methods presented in Section \ref{MAG} is shown in Table \ref{Table4}. The modified CF method described in Section \ref{DIS_B} is also included.


\begin{table}

\caption{Estimations of the Magnetic Field Strength.}
\label{Table4}

\begin{tabular}{lcc}
\hline

Method	&	B (mG)			\\
					
\hline

1: Paramagnetic alignment										&		4$-$82					\\
2: Thermal and magnetic equipartition							&		$>$6							\\
3: Chandrasekhar-Fermi method 								      &		52$^{+4}_{-8}$					\\
4: Modified Chandrasekhar-Fermi method$^{\mbox{\tiny a}}$				 	&		139$^{+11}_{-20}$		\\		

\hline
\end{tabular}\\
$^{\mbox{\tiny a}}$Modified CF Method is calculated as  $B' = B \times \beta^{-1/2}$, with $\beta = 0.15$ (Section \ref{DIS_B}).
\end{table}



\section{Discussion}
\label{DIS}


\subsection{Magnetic field in the torus of NGC 1068}
\label{DIS_B}

As discussed in Section \ref{ANA_INT_POL}, the measured polarisation arises from the passage of radiation through aligned dust grains in the torus of NGC 1068. As the grain alignment occurs with the long axes perpendicular to the magnetic field, a measured P.A. of polarisation parallel to the magnetic field is expected for dichroic absorption. We interpret the measured P.A. of polarisation at K$'$ in a 0$\farcs$5 (30 pc) aperture, PA$_{\mbox{\tiny K$'$}}$ = 127\degree $\pm$ 2\degree, as the orientation of the magnetic field in the torus of NGC 1068. We took the orientation of the torus to be PA$_{\mbox{\tiny torus}} \sim -$42\degree~(measured anti-clockwise, for direct comparison with our PA$_{\mbox{\tiny K$'$}}$, then PA$_{\mbox{\tiny torus}} \sim -$42\degree~+ 180\degree = 138\degree) from \citet{Raban2009}, and we found that the PA$_{\mbox{\tiny torus}} \sim$ PA$_{\mbox{\tiny K$'$}}$. The constant component of the magnetic field is thus aligned with the torus axis. Our result implies that our measured P.A. of polarisation gives us direct information on the existence and geometry of the torus. Specifically, the polarisation signature of the parsec-scale structure, identified as the obscuring torus, is observed through NIR polarisation and direct spatial comparison between IR interferometric and IR polarimetric observations can be made. Based on these results, the misalignment of $\sim$49\degree~between the measured P.A. of polarisation and the radio jet axis (Section \ref{RES}, Figure \ref{fig2}) can be interpreted as the misalignment between the torus (assuming that the torus axis is in the plane of the sky)  and the radio jet axis. Thus, physical components far below the spatial resolution of the NIR polarimetric observations can be estimated.  

As noted in Section \ref{MAG_M3}, we assumed the magnetic field as a mix of a turbulent and a constant component, where the turbulent component can be in any orientation. To infer the relative contribution of both components, we used the method presented by \citet{Hildebrand2009}. These authors showed that the ratio of turbulent to constant components of the magnetic field strength, $\langle B^{2}_{\mbox{\tiny t}}\rangle^{1/2} / B_{\tiny o}$, is a function of the dispersion of the plane of polarisation, $\alpha$, given by:  

\begin{equation}
\frac{\langle B^{2}_{\mbox{\tiny t}}\rangle^{1/2}}{B_{\tiny o}} = \frac{\alpha}{\sqrt{2-\alpha^2}}
\end{equation}

Using the dispersion of the plane of polarisation $\alpha$ = 7\degree~(0.1221 radians) from Section \ref{MAG_M3}, the ratio of turbulent to constant components of the magnetic field strength is  $\langle B^{2}_{\mbox{t}} \rangle^{1/2} / B_{0}$ = 0.09. Thus, the constant component, with a  toroidal geometry, is dominant in the torus of NGC 1068.

The misalignments of the structures in the inner few parsecs of NGC 1068 can produce local inhomogeneities in the magnetic field. These inhomogeneities produce an underestimation of the magnetic field strength measured through the paramagnetic alignment (Section \ref{MAG_M1}). This underestimation of the magnetic field strength can explain the similarity of the paramagnetic alignment and the  thermal and magnetic relaxation time equipartition. Thus, a higher magnetic field strength should be expected. The CF method assumed thermal and magnetic pressure equipartition. \citet{KB2003} showed that the equipartition is not satisfied in all regions of the molecular clouds and the CF method must be refined/modified in such regions. They defined the thermal to magnetic pressure ratio, $\beta$, as:

\begin{equation}
\beta = \frac{c_{\mbox{\tiny s}}^{2}}{V_{\mbox{\tiny A}}^{2}} = \frac{\rho c_{\mbox{\tiny s}}^{2}}{(B^{2}/4\pi)}
\label{beta}
\end{equation}

\noindent
where $c_{\mbox{\tiny s}}$ is the sound speed defined as $c_{\mbox{\tiny s}} = \sqrt{kT/m}$, $k$ is the Boltzmann constant, $T$ is the gas temperature, and $m$ is the mean molecular mass. $V_{\mbox{\tiny A}}$ is the Alfv\'en speed, defined as $V_{\mbox{\tiny A}}$ = $B/\sqrt{4\pi\rho}$, $\rho$ is the cloud density, and $B$ is the magnetic field strength. 

The dimensionless $\beta$ parameter is typically used in MHD simulations \citep[i.e.][]{OSG2001,KB2003} to account for the magnetization of the molecular clouds. Three different situations can be found: (1) $\beta \gg 1$ is considered as low magnetization, (2) $\beta \sim$ 1 represents the equipartition between dynamic and magnetic pressure (CF method condition), and (3) $\beta \sim$ 0 is considered as high magnetization. \citet{KB2003} found that the dynamic and magnetic pressure ratio is in approximate equipartition, $\beta \sim$ 1, in the regions containing most of the mass of the molecular clouds. However, $\beta <$ 1 in the low-density outer regions of the clouds.

Using the magnetic field strength, $B$ = 52$^{+4}_{-8}$  mG from the CF method, and the cloud density from Section \ref{MAG_M3}, we find $\beta$ $\sim$ 0.15. This value suggest that the clumps in the torus of NGC 1068 are in a highly magnetic environment. Another implication of $\beta \sim$ 0.15 is that the polarisation is arising from the low-density regions of the clumps. This physical condition agrees with the conditions of the paramagnetic alignment mechanism (Section \ref{MAG_M2}). Thus, the magnetic field strength estimated through the CF method should be refined using the $\beta$ parameter. The modified CF method is given by $B'$ = $B$ $\times$ $\beta^{-1/2}$ \citep{OSG2001}, which yields a magnetic field strength $B'$ = 139$^{+11}_{-20}$ mG (Table \ref{Table4}).

\begin{figure}
\includegraphics[angle=0,trim=0cm 0cm 0cm 0cm,scale=.23]{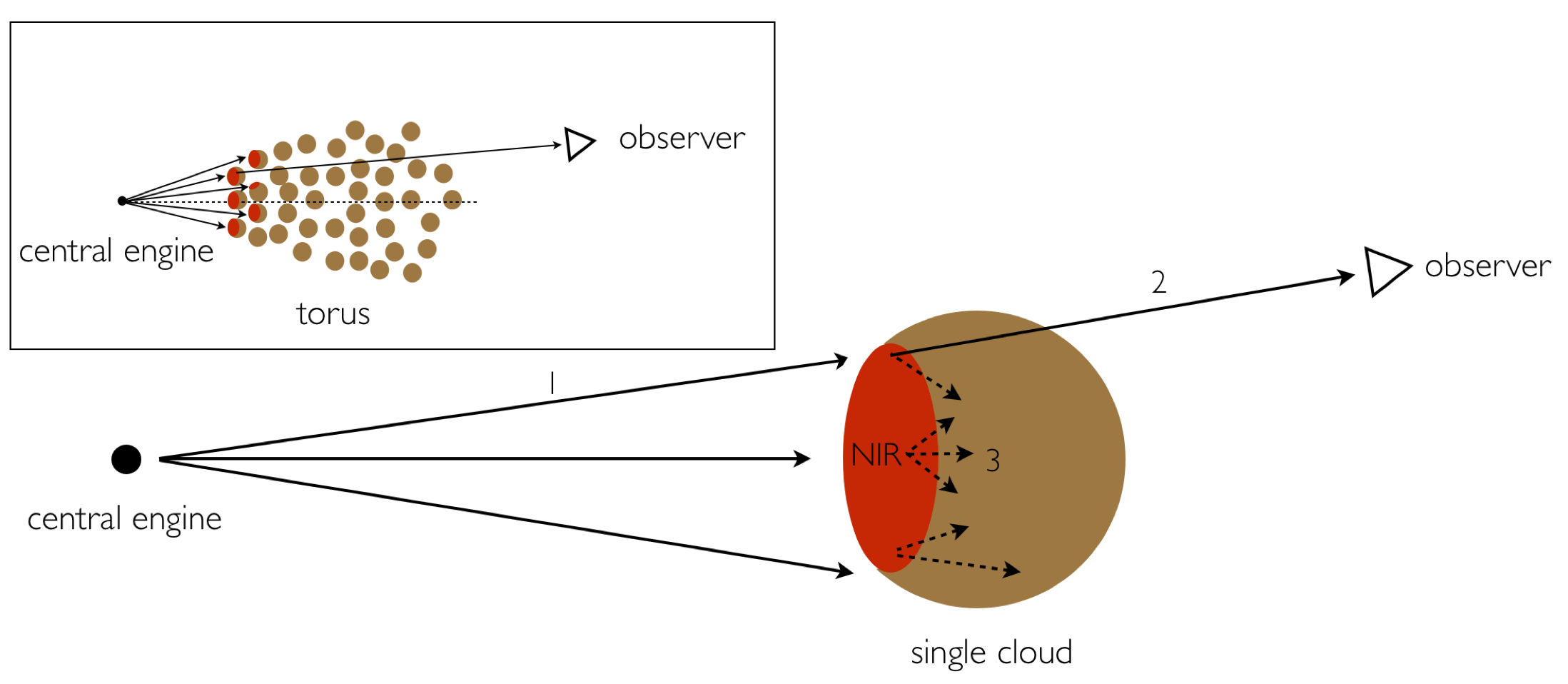}
\caption{Sketch of a single cloud and its location within the torus (box).  (1) The NIR total flux comes from the directly illuminated clumps located above the equational plane affording a direct view into our LOS, i.e., the surface of the clumps that are directly illuminated by the central engine \citep{HPB2008}, (2) the NIR radiation passes through the low-density outer layers of the clumps, and (3) NIR polarisation arises from the passage of radiation through the aligned dust grains in the clumps (Section \ref{DIS_B}).}
\label{fig4}
\end{figure}

We interpret these results as: (1) the NIR total flux comes from the directly illuminated clumps located above the equatorial plane affording a direct view into our LOS, i.e., the surface of the clumps that are directly illuminated by the central engine \citep[e.g.][]{HPB2008,HS2012,Stalevski2012}, (2) the NIR radiation passes through the low-density outer layers of the clumps, and (3) NIR polarisation arises from the passage of radiation through the aligned dust grains in the clumps. These dust grains are aligned by a global toroidal magnetic field in the torus generated by the accretion disc. The clumps located above the equatorial plane, those clumps in (1), have higher likelihood of escaping photons along our LOS than those located close to the equatorial plane.  A sketch is shown in Figure \ref{fig4}.


\subsection{Hydromagnetic wind}

Within the MHD framework \citep[e.g.][]{BP1982,EBS1992,Contopoulos1994,K1999}, the magnetic field plays an important role in the dynamics of the clumps in the torus. The inflow/outflow mass rates for those clumps showing NIR dichroic absorption are estimated assuming a hydromagnetic wind model. The mass accretion rate as a function of the magnetic field strength can be estimated through the thermal and magnetic pressure ratio. The thermal pressure on the midplane \citep{FKR1992} is given by: 

\begin{equation}
  P_{\mbox{\tiny th}} = \rho c_{\mbox{\tiny s}}^{2} = \frac{GM_{\mbox{\tiny BH}}\dot{M}}{3 \pi \sqrt{2\pi} c_{\mbox{\tiny s}} r^{3} \alpha_{\mbox{\tiny s}} }
\label{P_th}
\end{equation}

\noindent
where $M_{\mbox{\tiny BH}}$ is the black hole mass, $\dot{M}$ is the mass accretion rate in units of M$_{\odot}$ yr$^{-1}$, $r$ is the distance from the black hole in units of parsecs, and $\alpha_{\mbox{\tiny s}}$ is the Shakura-Sunyaev \citep{SS1973} viscosity parameter. 

As described in Section \ref{DIS_B}, the thermal and magnetic pressure equipartition was defined as $\beta = P_{\mbox{\tiny th}}/P_{\mbox{\tiny B}}$, which can be used in Equation (\ref{P_th}) to estimate the mass accretion rate as a function of the magnetic field:

\begin{equation}
\dot{M} = \frac{3\sqrt{2\pi}c_{\mbox{\tiny s}}r^{3} B^{2} \alpha_{\mbox{\tiny s}} \beta}{8GM_{\mbox{\tiny BH}}}
\label{M_dot}
\end{equation} 

The parameters $\alpha_{\mbox{\tiny s}}$ and $\beta$ are related through the MHD simulations prescription $\alpha_{\mbox{\tiny s}} = C/(1+\beta)$, where $C$ is a constant typically set in the range of 0.5$-$0.6 \citep{Narayan1998}. This relation is satisfied if magnetic fields provide the kinematic viscosity, which is the case in the magnetic environment in the clumps, $\beta \sim$ 0.15, estimated in Section \ref{DIS_B}. In this context, the viscosity parameter is estimated to be $\alpha_{\mbox{\tiny s}}$ = 0.48 $\pm$ 0.05. We use the magnetic field strength estimated to be in the range of $B =$ 4$-$139 mG (Section \ref{DIS}). We took the black hole mass of NGC 1068 $M_{\mbox{\tiny BH}}$ = $(8.0\pm0.3) \times 10^{6} $~M$_{\odot}$ from \citet{LB03}. These authors fitted a self-gravitating accretion disc model to the observed non-Keplerian rotational curve of the maser disc to derive the black hole mass in NGC 1068.  From the physical interpretation presented in Section $\ref{DIS_B}$, the NIR emitting region arises from the directly illuminated faces of the clumps located as  $r_{\mbox{\tiny sub}} =$ 0.4 pc for dust grains at 1500 K (Section \ref{MAG_cond}).

The clumps with measured NIR dichroism were assumed to be above the equatorial plane (Figure \ref{fig4}), these clumps are located in the inner edge of the torus, with a vertical heigh of $H  = r_{\mbox{\mbox{\tiny sub}}}\sin{\sigma} = 0.17^{+0.04}_{-0.02}$ pc. As a first approximation, the balance of magnetic and thermal pressure on the midplane allow us to estimate upper limits of the mass accretion/outflow rates. Based on these conditions and using Equation (\ref{M_dot}), we estimate the upper limit of the mass accretion rate to be:

\begin{align}
\dot{M} \le 8 \times 10^{-3} \left(\frac{r}{0.4~\mbox{pc}}\right)^{3} \left(\frac{T}{1500~\mbox{K}}\right)^{1/2}\left(\frac{B}{139~\mbox{mG}}\right)^{2} \nonumber \\
\times \left(\frac{M_{\mbox{\tiny BH}}}{8.0\times10^{6}~\mbox{M}_{\odot}}\right)^{-1}~~\mbox{M}_{\odot}~\mbox{yr}^{-1}
\label{M_dot_normalized}
\end{align}

For the canonical values in Equation (\ref{M_dot_normalized}), the mass accretion rate is $\dot{M} \le 8 \times 10^{-3}$ M$_{\odot}$ yr$^{-1}$ at a distance of 0.4 pc from the central engine of NGC 1068. As we only have physical information of those clouds within the outflow where we measure NIR dichroism, the estimated inflow/outflow mass accretion rate represents only a fraction of the black hole mass accretion rate, $\dot{M} = 0.18$ M$_{\odot}$ yr$^{-1}$, calculated using the bolometric luminosity, L$_{\tiny{bol}}$ = $9.55~\times~10^{44}$ erg s$^{-1}$ \citep{WU2002}, of NGC1068. 

\citet{ES2006} showed that the mass outflow wind rate, $\dot{M}_{\mbox{\tiny w}}$, and mass accretion rate, $\dot{M}$, can be related as: 

\begin{align}
\dot{M}_{\mbox{\tiny w}} \le 0.14 \left(\frac{T}{1500~\mbox{K}} \right)^{-2.6} \left(\frac{N_{\mbox{\tiny H}}}{10^{23}~\mbox{cm$^{-2}$}}\right) \nonumber \\
\times \left(\frac{v}{100~\mbox{km s$^{-1}$}}\right) I_{1} (\epsilon \dot{M})^{1/2}~~\mbox{M}_{\odot}~\mbox{yr}^{-1}
\end{align}
\noindent
where  $N_{\mbox{\tiny H}}$ is the total column density normalised to 10$^{23}$ cm$^{-2}$, $v$ is the velocity at the inner radius of the torus normalised to 100 km s$^{-1}$ based on H$_{2}$O maser VLBI observations of NGC 1068 at $\sim$0.4 pc by \citet{Greenhill1996}, $\epsilon$ is the accretion efficiency at the torus radii, and $I_{1}$ is an unknown factor of order unity \citep[Section 6.1 in][]{Nenkova2008b}.

We took the total column density, $N_{\mbox{\tiny H}} = 1.41^{+0.22}_{-0.37}$ $\times$ 10$^{24}$ cm$^{-2}$ (Section \ref{MAG_cond}), the velocity of 100 km s$^{-1}$, the accretion efficiency at the torus inner radius of $\epsilon = 0.01$, typically used at the torus scales by \citet{EBS1992,ES2006}, and we obtain an upper-limit of the mass outflow rate of $\dot{M}_{\mbox{\tiny w}} \le 0.17$  M$_{\odot}$ yr$^{-1}$ at a distance of 0.4 pc from the central engine of NGC 1068. 

The torus outflow timescale is  $t_{\mbox{\tiny w}} = M_{\mbox{\tiny torus}}/\dot{M}_{\mbox{\tiny w}}$. $M_{\mbox{\tiny torus}}$ is the torus mass given by $M_{\mbox{\tiny torus}} = m_{\mbox{\tiny p}}N_{\mbox{\tiny H}}\int{N_{\mbox{\tiny cl}}dV} \sim 10^{3} N_{\mbox{\tiny H},23}L_{45}Y$ M$_{\odot}$ \citep{ES2006}, where $L_{45}$ is the bolometric luminosity normalised to 10$^{45}$ erg s$^{-1}$, and $Y$ is the torus radial thickness. We took the bolometric luminosity to be L$_{\mbox{\tiny bol}} = 9.55 \times 10^{44}$ erg s$^{-1}$ \citep{WU2002}, and $Y = 6^{+2}_{-1}$ (Section \ref{MAG_cond}), and we estimate the mass of the torus to be $M_{\mbox{\tiny torus}} =  6.73^{+1.08}_{-1.74} \times 10^{4}$ M$_{\odot}$. This value is within a factor of ~three lower than $M_{\mbox{\tiny torus}} = 2.4 \times 10^{5}$ M$_{\odot}$ using radio observations by \citet{Gallimore1996} and $M_{torus} = 2.1(\pm1.2) \times 10^{5}$ M$_{\odot}$ using mm-observations with ALMA by \citet{GB2014}. The difference in torus mass can be interpreted as the \textsc{Clumpy} torus fitting to the IR (1$-$20 \um) SED is only accounting for those clouds emitting in the IR, where a more compact torus would be detected. The outflow timescale is estimated to be $t_{\mbox{\tiny w}} \geq 10^{5}$ yr. If we assume a torus rotating around the accretion disc with a typical Keplerian orbit of $t_{\mbox{\tiny K}} = 3 \times 10^{4} M_{\mbox{\tiny BH},7}^{-1/2} r^{3/2}_{\mbox{\tiny pc}}$ yr \citep{ES2006}, the timescale to complete a Keplerian orbit at 0.4 pc from the central engine is $t_{\mbox{\tiny K}} \sim 10^{3}$ yr. Thus, the obscuring structure of NGC 1068, generated by the outflowing wind, can be created in $\geq$100 Keplerian orbits, yielding a rotational velocity $\leq$1228 km s$^{-1}$. These results suggest that the outflowing wind, generated by the accretion disc's magnetic field, can rapidly create an dusty obscuring structure around the central engine of NGC 1068. Recent time-resolving 3D hydrodynamical models of AGN \citep{Schartmann2014} found that the radiation of the central engine powers a loss of mass with a rate of 0.1 M$_{\odot}$ yr$^{-1}$, with similar inflow mass rates from large scales in order to keep the dusty structure lifetime for a period of $\sim$10 Myr. In addition, some models assuming gas accretion flows, such as adiabatic inflow-outflow solutions (ADIOS, \citet{BB1999}) and convection-dominated accretion flows (CDAF, \citet{Narayan2000}), suggest that only a small fraction of the accreted matter at the outer radius of the inflow contributes to the mass accretion rate at the black hole. This can be explained by turbulence and strong mass lost. In this scheme, the accreting efficiency is low ($\epsilon \sim 0.01$) at the outer radius of the inflow, favoring the outflow wind. This scenario should be examined in a sample of AGNs to find general and/or extraordinary properties of the outflowing wind model.


\section{Conclusions}
\label{CON}

We found that the existence and evolution of the torus in NGC 1068 can be explained through the hydromagnetic outflowing model of AGNs. Assuming \textsc{Clumpy} torus models, the polarisation of NGC 1068 in the K$'$ filter is most likely arising from radiation of the directly illuminated inner-facing clumps of the torus passing through the magnetically aligned dust grains located in the low-density outer regions of the clumps. We found that the constant component of the magnetic field in the plane of the sky is dominant and responsible for the dust grain alignment in the torus with a strength in the range of 4$-$139 mG. We presented a direct comparison between the P.A. of the NIR polarimetric and IR interferometric observations of NGC 1068. Specifically, a toroidal geometry is the most likely configuration of the magnetic field in the torus. Adopting the estimated magnetic field configuration, we find a mass outflow rate of $\le$0.17 M$_{\odot}$ yr$^{-1}$ at a distance of 0.4 pc from the central engine. At this rate, the obscuring structure around the central engine can be created in a timescale of $\geq$10$^{5}$ yr with a rotational velocity of $\leq$1228 km s$^{-1}$. We conclude that the origin, evolution and kinematics of the dusty environment obscuring the central engine of NGC 1068 can be explained by a hydromagnetic outflow wind. Further NIR polarimetric observations of a sample of AGNs are required to refine and/or modify this approach and find general and/or extraordinary magnetic properties in the torus. In addition, MIR polarimetric observations are essential to test the effects of dichroism at longer wavelengths. Also, mm-polarimetric observations with ALMA will allow us to refine intrinsic properties of the torus, i.e. dust density, grain sizes, temperature, used to estimate the magnetic field strength through the different approaches presented in this work, as well as, to test the MHD outflow wind model for those clumps at the outer edge of the torus.

\section*{Acknowledgments}

The authors would like to thank the anonymous referee for the helpful comments. It is a pleasure to acknowledge discussion with R. Antonucci, T. D\'iaz-Santos, M. Imanishi, M. Kishimoto and M. Stalevski. This work is based on observations made with the 6.5-m MMT, Arizona, USA. E.L.R and C.P. acknowledge support from the University of Texas at San Antonio. C.P. acknowledges support from NSF-0904421 grant. C.P. and T.J.J acknowledge support from NSF-0704095 grant. AA-H acknowledges financial support from the Spanish Plan Nacional de Astronom\'ia y Astrofis\'ica under grant AYA2012-31447.  N.A.L. and R.E.M. are supported by the Gemini Observatory, which is operated by the Association of Universities for Research in Astronomy, Inc., on behalf of the international Gemini partnership of  Argentina, Australia, Brazil, Canada, Chile, and the United States of America. R.N. acknowledges support by FONDECYT grant No. 3140436. C.R.A. is supported by a Marie Curie Intra European Fellowship within the 7th European Community Framework Programme (PIEF-GA-2012-327934).

\end{document}